\documentclass[%
 aip,
 jmp,%
 amsmath,amssymb,
 reprint,%
] {revtex4-1}

\usepackage{graphicx}
\usepackage{dcolumn}
\usepackage{bm}

\begin{document}

\preprint{AIP/123-QED}

\title{Tendency to occupy a statistically dominant spatial state
of the flow as a driving force for turbulent transition}

\author{Sergei F. Chekmarev}
\noaffiliation
\affiliation{Institute of Thermophysics, SB RAS, 630090 Novosibirsk, Russia, and\\
Department of Physics, Novosibirsk State University, 630090 Novosibirsk, Russia}

\email{chekmarev@itp.nsc.ru}

\date{\today}

\begin{abstract}
\label{intro} The transition from laminar to turbulent fluid motion occurring at large
Reynolds numbers is generally associated with the instability of the laminar flow. On the
other hand, since the turbulent flow characteristically appears in the form of spatially
localized structures (e.g., eddies) filling the flow field, a tendency to occupy such a
structured state of the flow cannot be ruled out as a driving force for turbulent
transition. To examine this possibility, we propose a simple analytical model that treats
the flow as a collection of localized spatial structures, each of which consists of
elementary cells in which the behavior of the particles (atoms or molecules) is
uncorrelated. This allows us to introduce the Reynolds number, associating it with the
ratio between the total phase volume for the system and that for the elementary cell.
Using the principle of maximum entropy to calculate the most probable size distribution
of the localized structures, we show that as the Reynolds number increases, the
elementary cells group into the localized structures, which successfully explains
turbulent transition and some other general properties of turbulent flows. An important
feature of the present model is that a bridge between the spatial-statistical description
of the flow and hydrodynamic equations is established. We show that the basic assumptions
underlying the model, i.e. that the particles are indistinguishable and elementary
volumes of phase space exist in which the state of the particles is uncertain, are
involved in the derivation of the Navier-Stokes equation. Taking into account that the
model captures essential features of turbulent flows, this suggests that the driving
force for the turbulent transition is basically the same as in the present model, i.e.
the tendency of the system to occupy a statistically dominant state plays a key role. The
instability of the flow at high Reynolds numbers can then be a mechanism to initiate
structural rearrangement of the flow to find this state.
\end{abstract}

\pacs{47.27.Ak, 47.27Cn, 47.27.eb}
\keywords{fluid flow, transition to turbulence, driving force, statistical
model, maximum entropy principle, Navier-Stokes equation}
\maketitle

\begin{quotation}
As the flow velocity (or, more generally, the Reynolds number) increases, the fluid
motion becomes increasingly irregular and complex, what is known as laminar-to-turbulent
transition. The question "Why does this happen?", and a more challenging question "Is
there a target state to achieve?", are still under thorough discussion. One possible
viewpoint is that according to the general principles of statistical physics, the
observed flow presents a macrostate of the system that is obtained by the maximum number
of microstates available at given determining parameters, i.e. the maximum entropy.
Therefore, at different Reynolds numbers the observed macrostates may be different,
ranging from laminar to turbulent state. The idea to use the maximum entropy principle is
not new; for inviscid fluids, various variants of spatial organization of the flow and
counting of the microstates have been considered. Here, we extend this approach to
viscous fluids. The flow is represented by a collection of localized spatial structures,
each of which consists of elementary cells in which the behavior of particles (atoms or
molecules) is uncorrelated. This allows us to introduce the Reynolds number as a
determining parameter and throw a bridge from this spatial-statistical description of the
flow to hydrodynamic (Navier-Stokes) equations. The proposed model successfully predicts
transition to turbulence as the Reynolds number is increased and suggests that the
tendency of the system to occupy a statistically dominant state plays a basic role in
this phenomenon. The instability of the laminar flow can then be a mechanism to initiate
the structural rearrangement of the flow to find the dominant state.
\end{quotation}

\section{Introduction}
History of turbulence is long and complex
\cite{MoninYaglom,LandauLifshitz_fluid_mech,Frisch,Lesieur}, going back to seminal works
of Reynolds \cite{Reynolds1883}, Richardson \cite{Richardson1922} and Kolmogorov
\cite{Kolmogorov41a}. Originated to describe a complex and unpredictable motion of a
fluid, the concept of turbulence has become common to many fields of science; examples
are the behavior of waves on fluid surfaces and in plasma \cite{ZakharovLvovFalkovich},
dynamics of tangles of quantized vortices in superfluids
\cite{Vinen06,SkrbekSreenivasan12}, cascades in financial markets
\cite{financial_turbulence}, and dynamics of protein folding \cite{KalginChekmarev11}.
Although similar in some features, such as the presence of coherent flow structures and
cascade dynamics, the turbulent phenomena are very diverse. Even in the case of classical
fluid, turbulent flows have essentially different properties, e.g., pipe, jet or free
convective flows. Turbulence can vary from isotropic (e.g., grid turbulence
\cite{FrenkielKlebanoffHuang79}) to inherently anisotropic (shear turbulent flows, both
free and wall-bounded \cite{Lesieur}), the energy cascades between eddies with different
length scales can be opposite directed \cite {BoffettaEcke12} (as the direct cascades in
three-dimensional turbulence \cite{Richardson1922} and inverse cascades in
two-dimensional turbulence \cite{Kraichnan67}), etc. Still more complex is the transition
to turbulent motion, because it generally depends on how turbulence is triggered (the
inlet conditions for pipe flows \cite{Reynolds1883,Mullin11}, the interaction of
different instability modes \cite{Kachanov94} and free-stream perturbations
\cite{SaricReedKerschen02} for boundary layers, etc.). The linear stability analysis
\cite{Drazin02} of the canonical flows, such as the Taylor-Couette flow in the gap
between two rotating cylinders, and the Poiseuille flows in pipe and between two plates
(see, e.g., Landau and Lifshitz \cite{LandauLifshitz_fluid_mech}), leads to predictions
that either match laboratory experiments (the Taylor-Couette flow), or do not match (the
pipe flow), or match to some degree (the plane Poiseuille flow)
\cite{TrefethenTrefethenReddyDriscoll93,EckhardtSchneiderHofWesterweel07}. To have more
realistic results, e.g., for pipe flow, a finite disturbance of the flow is required to
be taken into account \cite{Ben-DavCohen07}.

With clear understanding that an analytical theory that would be able to describe
turbulence in all its complexity, and turbulent transition in particular, is hardly
possible, experimental and computer simulation studies have been put forward and made a
great progress during last decades \cite{Lesieur}, particularly, the studies based on new
techniques, such as the Particle Image Velocimetry \cite{BernardWallace02} and Direct
Numerical Simulation \cite{MoinMahesh98} methods, respectively. Nevertheless, the general
question why, in principle, laminar motion becomes unfavorable at high Reynolds numbers
and changes to turbulent motion has not been completely answered. The stability analysis
of solutions of Navier-Stokes equations helps to answer to the first part of this
question, but it does not explain why the resulting flow, instead of staying disordered,
becomes partly organized in the form of coherent structures (Hussain \cite{Hussain86}).
The answer to this question was sought in terms of the dynamical system theory, whose
goal was to reach qualitative understanding of turbulent transition through model
nonlinear equations (see Eckmann \cite{Eckmann81} and Screenivasan \cite{Screenivasan99}
for review). The milestones are as follows. Landau \cite{LandauLifshitz_fluid_mech} and
Hopf \cite{Hopf48} proposed that the transition to turbulence presents a superposition of
successive bifurcations to smaller and smaller scales of fluid motions, which results in
complex quasiperiodic motion. Later, Lorenz \cite{Lorenz63} found that a solution of
deterministic equations (mimicking atmospheric fluid motion) is very sensitive to initial
conditions and ends on a subdomain of phase space that was later called "strange
attractor". Ruelle and Takens \cite{RuelleTakens71}, who coined this term, have shown
that a small perturbation of the quasi-periodic motion gives the flow with a strange
attractor. Feigenbaum \cite{Feigenbaum78} considered the transition to turbulence as a
period-doubling subharmonic sequence, and Pomeau and Manneville \cite{PomeauManneville80}
proposed that it involves intermittencies.

Another possible approach to understanding of turbulent transition, which is statistical
rather than dynamical, is that the turbulent state presents a macrostate obtained by
maximum number of elementary states available at given determining parameters, i.e. the
state corresponding to the maximum entropy. Then a tendency to occupy such a dominant
state can be considered to be a driving force for turbulent transition. This viewpoint is
supported by the fact that the turbulent state characteristically appears in the form of
collection of spatially localized (coherent) structures \cite{Hussain86}; for isotropic
turbulence such structures are known as vortex "worms" \cite{JimenezWraySaffmanRogallo93}
and for shear flows as filaments \cite{CadotDouadyCouder95}. The idea that the principle
of maximum entropy can be used to determine turbulent state goes back to Burgers
\cite{Burgers48} and Onsager \cite{Onsager49} (see Eyink and Sreenivasan
\cite{EyinkSreenivasan06} for review); to count the number of states, Burgers divided the
phase space of a fluid flow into discrete cells, and Onsager considered turbulent state
as a set of discrete (point) vortices. This approach has been used and developed by many
researchers; in particular, in application to stellar systems, Lynden-Bell
\cite{Lynden-Bell67} divided phase space into macrocells composed of microcells,
Kraichnan and Montgomery \cite{KraichnanMontgomery80} used a set of discrete vortices to
describe two-dimensional turbulence, and, recently, Jung et al.
\cite{JungMorrisonSwinney06} employed the Lynden-Bell approach for the same purpose. In
these studies the fluid was assumed to be an ideal fluid, i.e. with no viscosity;
accordingly, the Reynolds number was not introduced, and the evolution of the state of
the flow with the Reynolds number was not considered. To consider competition between
laminar and turbulent states, primarily in application to intermittency phenomena, Kaneko
\cite{Kaneko85} and Pomeau \cite{Pomeau86} proposed to view the flow as a mixture of two
spatially separated macroscopic phases that represent these states, i.e. to pass to a
"global" scale of characterization of laminar and turbulent states. Examples of such
turbulent states can be turbulent spots in plane Poiseuille
\cite{CarlsonWidnallPeeters82} and Couette \cite{TillmarkAlfredsson92} flows and "puffs"
\cite{EckhardtSchneiderHofWesterweel07, Avila_Hof11} in pipe flows. Using model equations
(coupled map lattices \cite{Kaneko85} and a reaction-diffusion equation applied to a
double-well potential \cite{Pomeau86}), it has been shown that depending on some control
parameter (mimicking the Reynolds number), one of the competing states can go from stable
to metastable or disappear \cite{Pomeau86} and a kind of "fully developed turbulence" can
be achieved due spatial coupling \cite{Kaneko85}. Such a characterization of turbulent
states on an overall level, i.e. without resolution of their (spatial) structure, has
been found useful both for developing theoretical models and for interpretation of the
results of numerical simulations and experiments. Some recent examples are the
two-variable model for the transition pipe flow (Barkley \cite{Barkley11}), the size
characterization of laminar domains embedded in turbulent flow (the simulation study of
the decay of turbulence in a plane Couette flow; Manneville \cite{Manneville09}), and the
interpretation of experimental results for pipe flow (Eckhardt et al.
\cite{EckhardtSchneiderHofWesterweel07} and Avila et al. \cite{Avila_Hof11}). In
particular, Eckhardt et al. considered the turbulent state as a chaotic saddle in state
space which consists of regions associated with laminar motion and coherent structures,
and Avila et al. studied the competition between the splitting and decay of puffs in pipe
flow, concluding that in contrast to the classical Landau-Ruelle-Takens theory
\cite{LandauLifshitz_fluid_mech,RuelleTakens71}, spatial proliferation of chaotic domains
is intrinsic to the nature of fluid turbulence.

In the present paper, we consider turbulent transition from a spatial-statistical
viewpoint somewhat similar to that in Refs. \onlinecite{Burgers48,Onsager49,
EyinkSreenivasan06, Lynden-Bell67,KraichnanMontgomery80,JungMorrisonSwinney06}, i.e
turbulent flow is represented by a collection of localized spatial structures. However,
in contrast to those works, the structures are assumed to consist of elementary cells in
which the behavior of the particles (atoms or molecules) is uncorrelated. To each such
cell there corresponds an elementary volume of phase space, in which the state of the
particles is uncertain (to avoid confusion, we will refer to the elements of physical
space as {\it cells}, and to the elements of phase space as {\it volumes}); for a gas
fluid, the velocity and linear scales of the volume are, respectively, molecular thermal
velocity and mean free path, whose product is the kinematic viscosity. In contrast to the
previous analytical models
\cite{Burgers48,Onsager49,EyinkSreenivasan06,Lynden-Bell67,KraichnanMontgomery80,
JungMorrisonSwinney06}, which were limited to consideration of {\it inviscid} fluids,
this definition of the elementary cell allows us to introduce the Reynolds number, i.e.
to extend the spatial-statistical description of the flow to {\it viscous} fluids. The
Reynolds number is determined as the cube root of the ratio of the total phase volume for
the system to that for the elementary cell. Calculating the most probable size
distributions of the localized structures and comparing their weights for different
Reynolds numbers, we show that the elementary cells cluster into the localized structures
as the Reynolds number increases. This successfully explains turbulent transition
occurring at large Reynolds numbers and some other characteristic properties of turbulent
flows, in particular, the bounds of the transition range depending on the Reynolds
number. An essential feature of the present model is that a connection between the "pure"
statistical description of the flow and hydrodynamic equations is established. It is
shown that the basic assumptions underlying the model, i.e. that the particles are
indistinguishable and the elementary volumes of phase space exist where the state of the
particles is uncertain, are involved in the derivation of the Navier-Stokes equation;
they can be said to play a hidden statistical content of the Navier-Stokes equation. This
suggests that the driving force for the turbulent transition described with the
hydrodynamic equations is basically the same as in the present model, i.e. the tendency
of the system to occupy a statistically dominant state plays a key role. The instability
of the flow can then be a mechanism to initiate the structural rearrangement of the flow
to find this state.

The paper is organized as follows. Section II describes the proposed statistical model
and introduces the Reynolds number. Section III presents the results. It discusses
turbulent transition and the bounds of the turbulent transition region (IIIA), some other
properties of turbulent flows predicted with the present model (IIIB), a different type
of the localized structures (IIIC), and the connection of the model with hydrodynamic
(Navier-Stokes) equations. Section IV presents the resulting conclusions.

\section{Model}
Let a fluid flow be represented by a system of $N$ identical particles (atoms or
molecules). Assume that the particles can form spatially localized structures of $N_i$
particles ($1 \le N_i \le N$), with the number of such ($i$) structures being $M_i$.
These structures can be viewed, in a broad sense \cite{Lesieur}, as coherent structures,
i.e. as the structures in which the constituting particles execute an overall concerted
motion on the structure scale, similar to what is observed, e.g., in the coherent
vortices \cite{CadotDouadyCouder95,JimenezWraySaffmanRogallo93}. Since the concerted
motion should break down at the microscale level, each $i$ structure can be divided into
$N_{i}/n$ elementary "cells" (each of $n$ particles), in which the behavior of the
particles is uncorrelated. Alternatively, this level can be associated with the
Kolmogorov dissipation scale (see Sect. IIIB below). To each such cell there corresponds
an elementary volume of phase space, in which the state of the particles is uncertain;
for a gas fluid, the velocity and linear scales of the volume are, respectively,
molecular thermal velocity and mean free path. Turbulent state is associated with the
structures that contain many elementary cells, i.e. $N_i/n \gg 1$, and laminar state with
those for which $N_i/n \sim 1$. Further assume that the interaction between the localized
structures is small so that the total kinetic energy of concerted motion in the
structures $E$ can be written as $E=\sum_{i}M_{i}E_{i}$, where $E_{i}$ is the kinetic
energy of concerted motion of the particles in $i$ structure. Every distribution of the
particles among the localized structures that obeys the conditions $\sum_{i}N_{i}M_{i}=N$
and $E=\sum_{i}M_{i}E_{i}$ presents a {\it microstate} of the system at given $N$ and
$E$. Taking into account that the particles are identical, so that the permutations of
the particles in the elementary cells, the permutations of the elementary cells in the
localized structures, and the permutations of the localized structures themselves do not
lead to new states, the total number of possible microstates can be estimated as for an
ideal-gas system \cite{LandauLifshitz_stat_phys}
$$\Gamma=\frac{N!}{\prod_{i}\left[(n!)^{N_{i}/n}(N_{i}/n)!\right]^{M_i}M_{i}!}$$
or, with the Stirling approximation $x!\approx (x/e)^x$ to be applicable
\begin{equation}\label{eq1}
 \Gamma=\frac{(N/n)^{N}e^{N/n}}{\prod_{i}(N_{i}/n)^{N_{i}M_{i}/n}(M_{i}/e)^{M_{i}}}
\end{equation}

With Eq. (\ref{eq1}), it is possible to calculate the most probable size distribution of
the structures, $\tilde{M}_{i}=\tilde{M}_{i}(N_{i})$, which maximizes the number of
states $\Gamma$ and, correspondingly, the entropy $S=\ln \Gamma$. It determines the
collection of localized structures of different size that is formed with a dominant
probability, i.e. the observable {\it macrostate} of the system. The most probable
distribution reduces the phase space of the system to a subspace of much lower dimension,
similar to how, according to the dynamical system theory, the phase space shrinks when
the system reaches the (strange) attractor in the transition to turbulence
\cite{LandauLifshitz_fluid_mech,RuelleTakens71}. Applying the Lagrange multiplier method
to the entropy functional, i.e. varying the expression $\sum_{i}\left[-(M_{i}N_{i}/n)\ln
(N_{i}/n)-M_{i}\ln(M_{i}/e)\right] +\alpha'\sum_{i}M_{i}N_{i}+\beta'\sum_{i}M_{i}E_{i}$
with respect to $M_{i}$  at two conservation conditions $\sum_{i}M_{i}N_{i}=N$ and
$\sum_{i}M_{i}E_{i}=E$, one obtains
\begin{equation}\label{eq2}
\tilde{M}_{i}=\left(\frac{N_{i}}{n}\right)^{-\frac{N_{i}}{n}}e^{\alpha' N_{i}+\beta'
E_{i}}
\end{equation}
where and $\alpha'$ and $\beta'$ are the constants dependent on $N$ and $E$ (the Lagrange
multipliers).

To proceed further, the dependence of $E_{i}$ on $N_{i}$ should be specified. To do this,
we will follow Kolmogorov \cite{Kolmogorov41a}, assuming that the velocity of concerted
motion of the particles in the structures varies with distance as
\begin{equation}\label{eq3}
v(r) \sim r^{h}
\end{equation}
where $h=1/3$. Although the Kolmogorov theory \cite{Kolmogorov41a} implies turbulence to
be locally homogeneous and isotropic, the range of its application is, in fact, wider;
e.g., as has been recently shown by Gioia and Chakraborty \cite {GioiaChakraborty06}, it
works well to describe turbulent friction in rough pipes. Strictly speaking, Eq.
(\ref{eq3}) assumes the inertial interval of scales $\eta \ll r \ll L$, where $\eta$ and
$L$ are the dissipation and external scales, respectively \cite{Kolmogorov41a}. However,
for rough estimates, the range of the validity of this equation can be extended to the
lower and upper bounds. The lower bound has been considered by Landau and Lifshitz
\cite{LandauLifshitz_fluid_mech}, who used $\eta$ as a point to match Eq. (\ref{eq3}) to
the corresponding dependence at $r \ll \eta$ (Ref. \onlinecite{small_scales}), where the
flow is regular and can be associated with laminar motion. We will show later (Sect.
IIIB) that the dissipation scale $\eta$ characterizes the linear size of the elementary
cells, and thus the structures of size $N_i/n \sim 1$ can represent laminar state. Also,
we will show there that the extension of the upper bound of variation of $r$ to the
external scale $L$ is quite acceptable as well, e.g., it leads to the correct dependence
of $\eta$ on the Reynolds number [Eq. (\ref{eq8})]. Therefore, we assume that Eq.
(\ref{eq3}) can be used throughout the entire range of variation of the structure size
($1 \le N_i/n \le N/n$).

With Eq. (\ref{eq3}), the kinetic energy per unit mass is $e(r) \sim r^{2h}$, and $E_{i}
\sim \int_{0}^{r_i} e(r)r^{2}dr \sim r_{i}^{2h+3} \sim N_{i}^{2h/3+1}$. Also, it is
convenient to pass from $N_{i}$ to the number of the elementary cells in the localized
structure $q_{i}=N_{i}/n$. With these changes, Eq. (\ref{eq2}) becomes
\begin{equation}\label{eq4}
\tilde{M}_{i}=q_{i}^{-q_{i}}e^{\alpha q_{i}+\beta q_{i}^{2h/3+1}}
\end{equation}
where $\alpha$ and $\beta$ are new constants that are determined from the equations of
conservation of the total number of particles and kinetic energy
\begin{equation}\label{eq5}
N/n=\sum_{i}\tilde{M}_{i}q_{i}
\end{equation}
and
\begin{equation}\label{eq6}
E/n^{2h/3+1}=\sum_{i}\tilde{M}_{i}q_{i}^{2h/3+1}
\end{equation}

Given the values of $n$, $N$ and $E$, the constants $\alpha$ and $\beta$ can be
calculated by generating a set of $q_{i}$ and solving Eqs. (\ref{eq5}) and (\ref{eq6})
with $\tilde{M}_{i}$ substituted from Eq. (\ref{eq4}). Another possibility is to vary
$\alpha$ and $\beta$ to obtain $N/n$ and $E/n^{2h/3+1}$ as functions of $\alpha$ and
$\beta$.

The parameter $N/n$ can be associated with the Reynolds number ${\mathrm
{Re}}_{L}=VL/\nu$, where $V$ and $L$ are, respectively, the velocity and linear scales
characterizing the system as a whole, and $\nu$ is the kinematic viscosity of the fluid.
Since the particles are identical, the $6N$-dimensional phase space is reduced to the
$6$-dimensional single-particle space. Then, the linear size of the elementary phase
space volume, in which the state of the particles is uncertain, can be taken as the
product of the characteristic velocity and length at which the state of the particles is
uncertain. For a gas fluid, it is the product of the molecular thermal velocity $c$ and
the mean free path $\lambda$ (Kusukawa \cite{Kusukawa51}). Then, the elementary phase
volume is $(c \lambda)^3$, with the number of particles in it being proportional to the
volume, $n \sim (c \lambda)^3$. Correspondingly, the total number of particles $N$ can be
considered to be proportional to the total phase volume for the system $(VL)^3$, i.e. $N
\sim (VL)^3$ (the density of the fluid is assumed to be constant). Taking into account
the kinetic theory expression for the gas viscosity $\nu \sim c \lambda$ (e.g., Chapman
and Cowling \cite{ChapmanCowling}), one obtains
\begin{equation}\label{eq7}
N/n \sim (VL/\nu)^3 ={\mathrm {Re}}_{L}^3
\end{equation}
A similar relation is valid for a liquid fluid. Here, the linear size of the elementary
volume can be taken to be $c^2 \tau$, where $c^2$ presents the fluctuations of the
(molecular) kinetic energy per unit mass, and $\tau$ is the mean residence time of the
molecule in a settled state. Since the viscosity of liquid is $\nu \sim c^2 \tau$ (e.g.,
Frenkel \cite{Frenkel}), we arrive at Eq. (\ref{eq7}) again. In what follows, we will
assume for simplicity that the coefficient of proportionality in Eq. (\ref{eq7}) is equal
to 1, i.e.
\begin{equation}\label{eq7a}
{\mathrm {Re}}_{L}=(N/n)^{1/3}
\end{equation}

To complete the description of the model, we note that the collection of localized
structures can be considered to consist of sets of structures that are not in
equilibrium. This is consistent with Eq. (\ref{eq1}), which is applicable for estimation
of entropy in nonequilibrium conditions \cite{LandauLifshitz_stat_phys}. Then, the set
the structures of size $N_i/n \gg 1$ could be viewed as a turbulent packet, such as the
"puffs" in pipe flows \cite{EckhardtSchneiderHofWesterweel07, Avila_Hof11} and turbulent
spots in plane Poiseuille \cite{CarlsonWidnallPeeters82} and Couette
\cite{TillmarkAlfredsson92} flows, and the rest of the structures (of size $N_i/n \sim
1$) as a laminar background, i.e. the intermittency phenomena could be considered.
However, the application of the maximum entropy approach to obtain the structure size
distribution [Eq. (\ref{eq2})] restricts the consideration to the equilibrium conditions,
and it is highly questionable if a description in terms of equilibrium statistics is
relevant to these phenomena, which are inherently nonequilibrium. Therefore, in the
present study, we will not deal with the intermittency phenomena.

\begin{figure}
\centering
\resizebox{0.8\columnwidth}{!}{ \includegraphics* {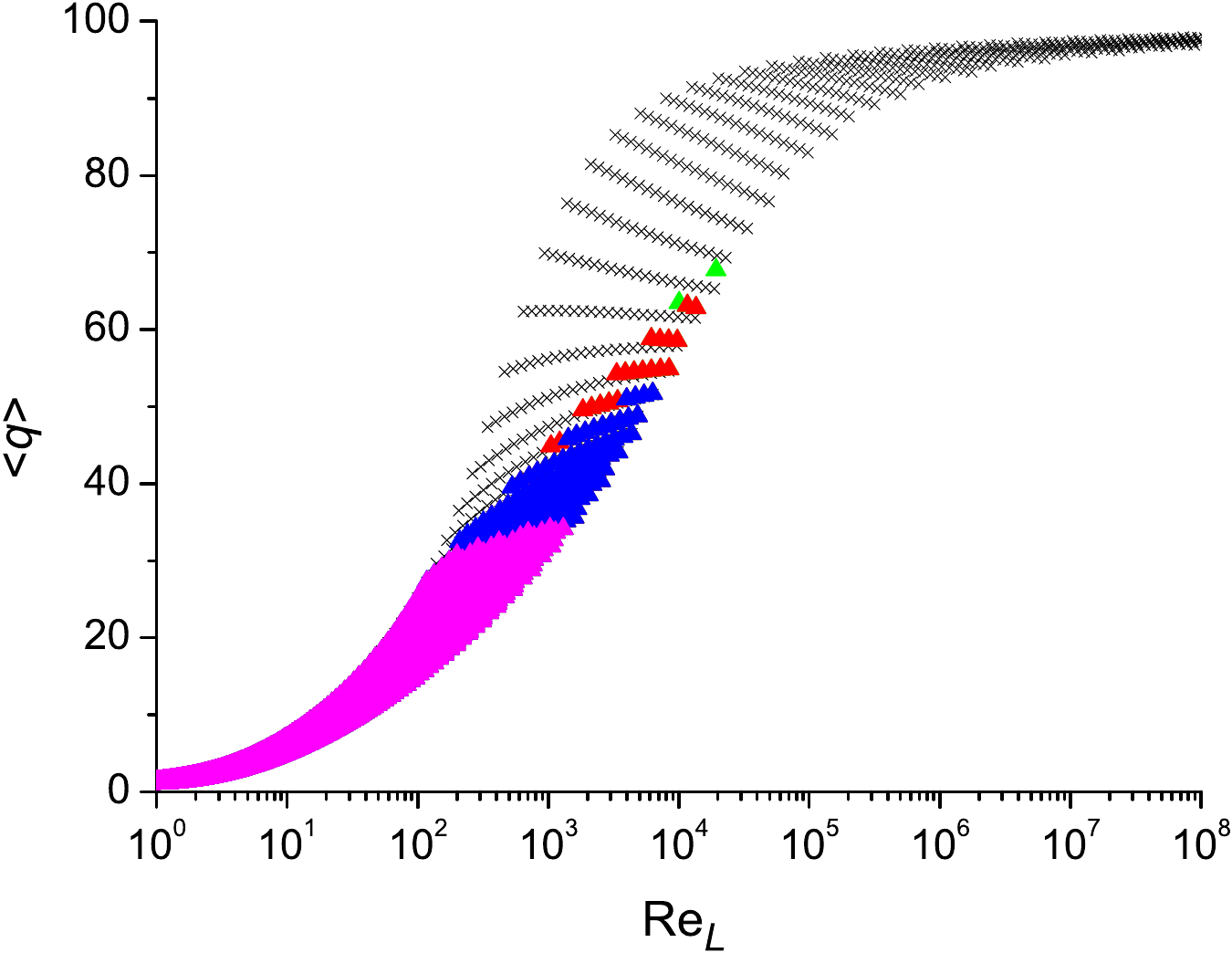}}%
\caption{Average number of the elementary cells in the localized structures $\langle
q\rangle$ as a function of $\mathrm{Re}_{L}$. The number of the elementary cells $q_i$
varies from 1 to $q_{\mathrm{max}}$, and parameters $\alpha$ and $\beta$ vary
independently from $\alpha_{\mathrm{min}}=\beta_{\mathrm{min}}=-1.5$ to
$\alpha_{\mathrm{max}}=\beta_{\mathrm{max}}=1.5$ with the interval 0.015. Crosses
represent all combinations of $\alpha$ and $\beta$ at $q_{\mathrm{max}}=100$, and the
triangles stand for the combinations which lead to the physically reasonable bell-shaped
distribution at different $q_{\mathrm{max}}$; the magenta, blue, red and green triangles
are for $q_{\mathrm{max}}=100, 200, 300$ and 400, respectively.}
\label{fig1}       
\end{figure}

\begin{figure}
\centering
\resizebox{0.8\columnwidth}{!}{ \includegraphics* {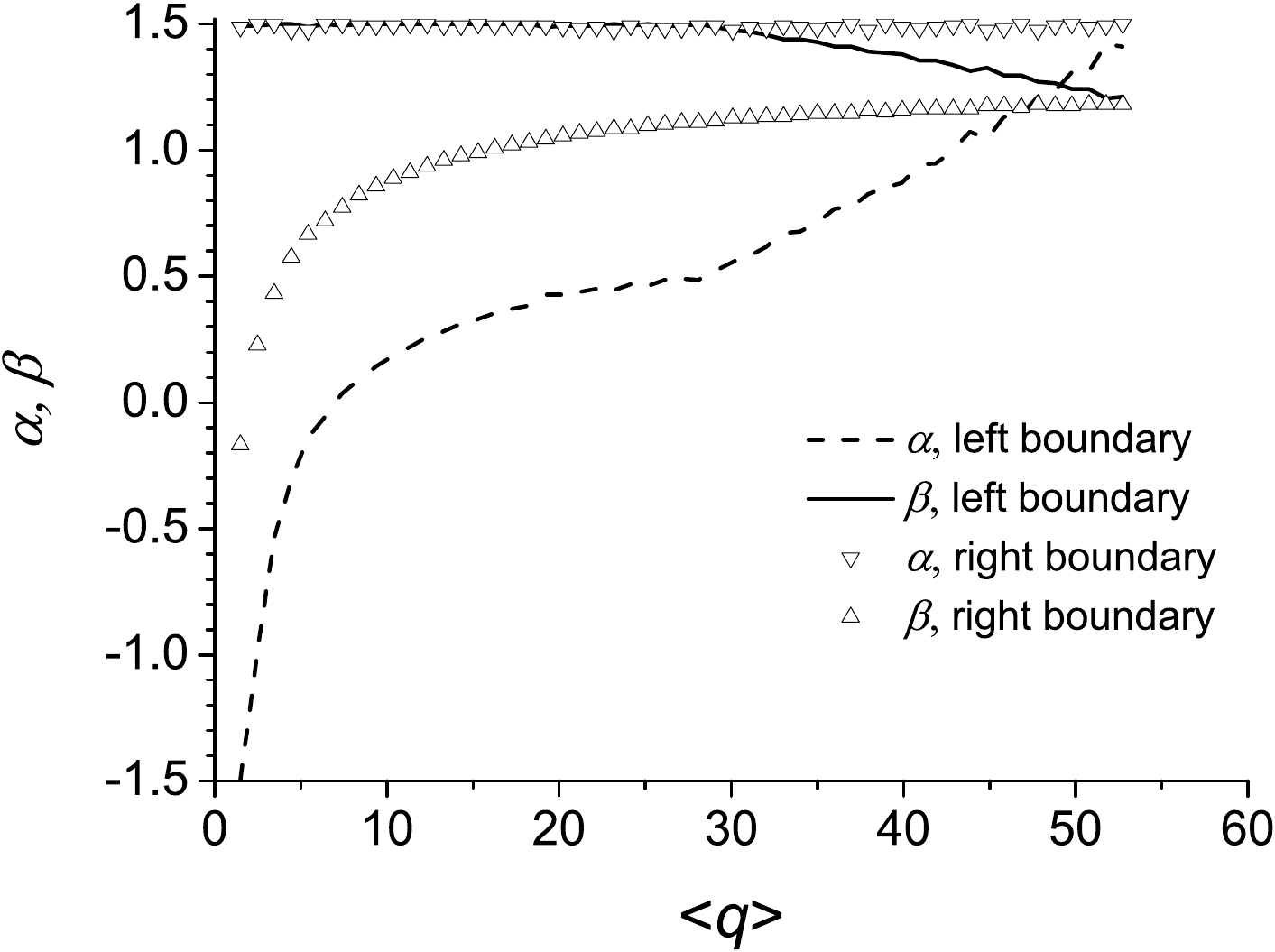}}%
\caption{The values of the parameters $\alpha$ and $\beta$ at the lower and upper
boundaries of the transition region of Fig. \ref{fig1}; $q_{\mathrm{max}}=200$.}
\label{fig2}       
\end{figure}

\begin{figure}
\centering
\resizebox{0.8\columnwidth}{!}{ \includegraphics* {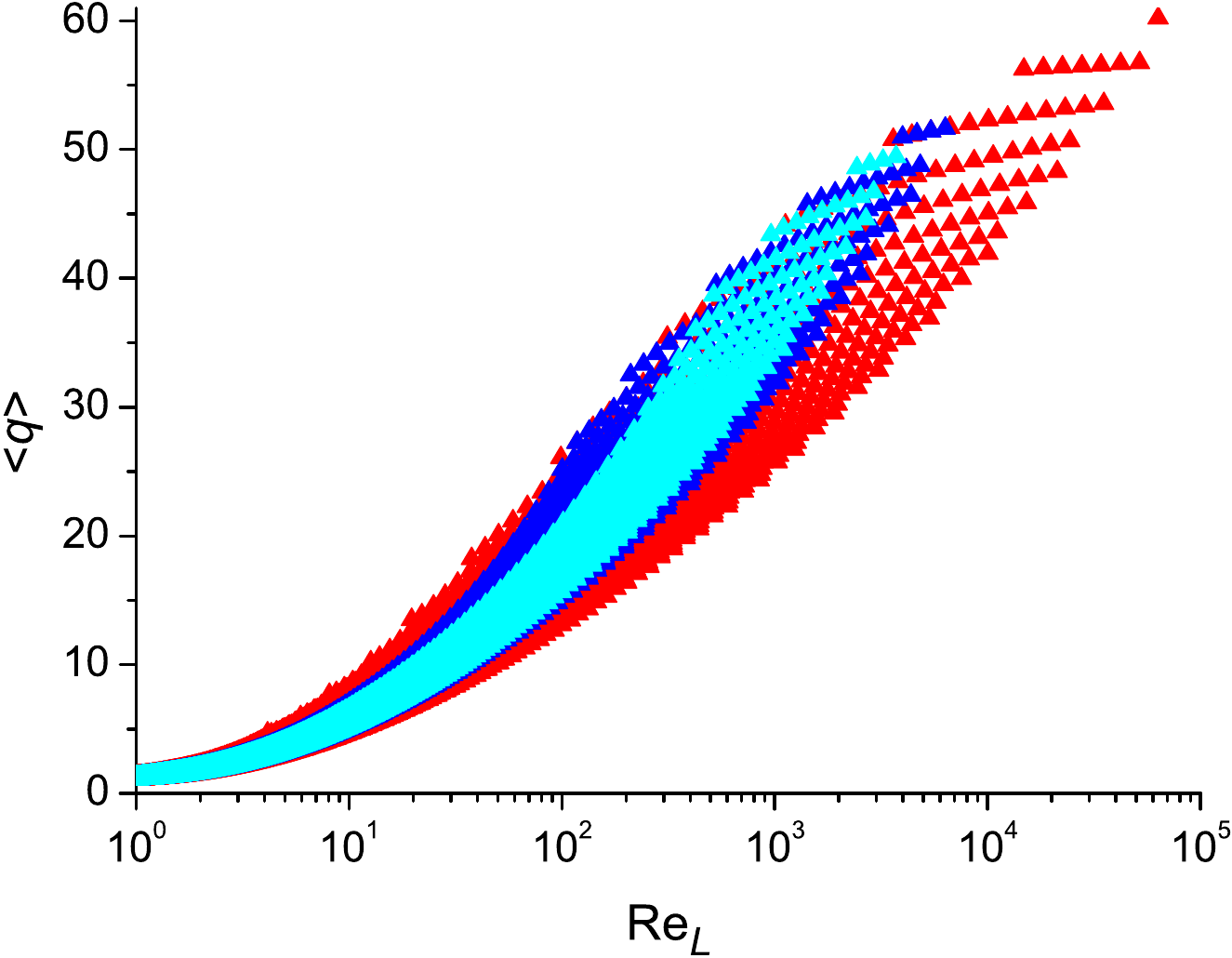}}%
\caption{Average number of the elementary cells in the localized structures $\langle q
\rangle$ as a function of $\mathrm{Re}_{L}$; $1 \leq q_{i} \leq 200$. The cyan triangles
are for $-1.3 \leq \alpha \leq 1.3$ and $-1.3 \leq \beta \leq 1.3$, the blue triangles
are for $-1.5 \leq \alpha \leq 1.5$ and $-1.5 \leq \beta \leq 1.5$ (as in Fig. 1), and
the red triangles are for $-2.0 \leq \alpha \leq 2.0$ and $-2.0 \leq \beta \leq 2.0$.}
\label{fig3}       
\end{figure}

\section{Results and Discussion}
\subsection{Turbulent Transition Region}
Figure \ref{fig1} shows a characteristic dependence of the average number of elementary
cells in the localized structures $\langle
q\rangle=\sum_{i}q_{i}\tilde{M}_{i}/\sum_{i}\tilde{M}_{i}$ on the Reynolds number
${\mathrm {Re}}_{L}$; the number of elementary cells $q_i$ varies from 1 to
$q_{\mathrm{max}}$. Points labelled with crosses represent all combinations of $\alpha$
and $\beta$, with these parameters varied independently within $\alpha_{\mathrm{min}}
\leq \alpha \leq \alpha_{\mathrm{max}}$ and $\beta_{\mathrm{min}} \leq \beta \leq
\beta_{\mathrm{max}}$ \cite{alpha_beta}. However, not every combination leads to a
physically reasonable bell-shaped distribution \cite{JimenezWraySaffmanRogallo93} (see
also Fig. \ref{fig5} below). In small fraction of the combinations (typically at large
values of $\beta$), $\tilde{M}_{i}$ does not vanish at $q \rightarrow q_{\mathrm{max}}$.
The fraction of such "wrong" combinations is generally within several percentages (in the
given case, below 2 percent). Triangles show the dependence $\langle q\rangle$ on
${\mathrm {Re}}_{L}$ with those "wrong" $\alpha / \beta$ combinations excluded. The
increase of $q_{\mathrm{max}}$ does not change the distribution, except that the range of
variation of $\langle q\rangle$ and ${\mathrm {Re}}_{L}$ is extended to larger values of
these quantities.

\begin{figure}
\centering
\resizebox{0.8\columnwidth}{!}{ \includegraphics* {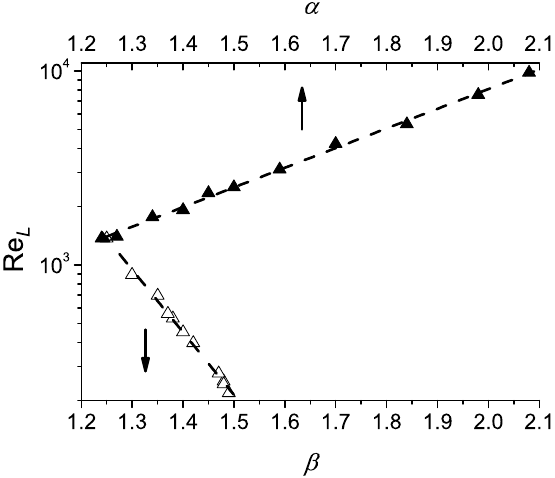}}%
\caption{The Reynolds number ${\mathrm {Re}}_{L}$ corresponding to the lower (open
triangles) and the upper (solid triangles) boundaries of the transition range as a
function of parameters $\beta$ and $\alpha$, respectively; $\langle q\rangle=40$. The
dashed lines are the best exponential fits, see the text.}
\label{fig4}       
\end{figure}

It is seen that as ${\mathrm {Re}}_{L}$ exceeds some characteristic value (${\mathrm
{Re}}_{L} \sim 10^2$ in Fig. \ref{fig1}), $\langle q\rangle$ rapidly increases, i.e. the
elementary cells group into localized structures, signaling that the flow becomes
turbulent at these Reynolds numbers. It is noteworthy that the dependence of $\langle q
\rangle$ on ${\mathrm {Re}}_{L}$ is not unique, i.e. the same values of $\langle
q\rangle$ are observed in a broad range of ${\mathrm {Re}}_{L}$. This indicates that the
turbulent state is not solely determined by the Reynolds number; rather, as is
well-known, it is flow specific, depending on the type of the flow, the inlet conditions,
including the magnitude of flow disturbance, and the flow environment \cite{Lesieur}. The
present statistical model is too simple to take these effects into account, but it offers
an estimate for the range of ${\mathrm {Re}}_{L}$ at which the turbulent motion can be
expected, i.e. the lower and upper bounds of the turbulent transition region. The lower
and upper boundaries of the manifold of the "correct" points in Fig. \ref{fig1} (labelled
with triangles) are determined, respectively, by the maximum values of $\beta$ and
$\alpha$ at which the correct points are obtained (Fig. \ref{fig2}). As these parameters
increase, the lower boundary shifts to smaller values of ${\mathrm {Re}}_{L}$, and the
upper boundary to larger values of ${\mathrm {Re}}_{L}$ (Fig. \ref{fig3}). More
specifically, at the lower boundary ${\mathrm {Re}}_{L} \sim \exp(-0.15 \beta\langle
q\rangle)$, and at the upper boundary ${\mathrm {Re}}_{L} \sim \exp(0.06 \alpha\langle
q\rangle)$. Figure \ref{fig4} shows these dependencies for $\langle q\rangle=40$, which
corresponds to approximately the maxima of the size distributions of the localized
structures (see Fig. \ref{fig5} and its discussion below). It is remarkable that while
the upper boundary monotonically shifts to larger values of $\mathrm{Re}_L$ as $\alpha$
increases, the lower boundary shifts to smaller values of $\mathrm{Re}_L$ until ${\mathrm
{Re}}_{L} \sim 10^2$ ($\beta^{*} \approx 1.5$) is reached; after that it "freezes" (Figs.
\ref{fig3} and \ref{fig4}). The model thus predicts that the flow should be laminar at
${\mathrm {Re}}_{L} < {\mathrm {Re}}_{L}^{\star}$ (${\mathrm {Re}}_{L}^{\star} \sim 10^2$
in the present model), and it can remain laminar up to very large values of ${\mathrm
{Re}}_{L}$.  This prediction is in good general agreement with the experimental and
simulation results \cite{LandauLifshitz_fluid_mech,Lesieur}. In particular, for the pipe
flow, Reynolds \cite{Reynolds1883} has found that the critical Re number varied from
$\mathrm{Re}_{L} \approx 2 \times 10^3$ to $\mathrm{Re}_{L} \approx 1.3 \times 10^4$
depending on the inlet conditions. The lower bound of the stability of the laminar flow,
which is observed at large disturbances of the inlet flow, has been confirmed in many
experimental and theoretical works (e.g., Darbyshire and Mullin
\cite{DarbyshireMullin95}, and Ben-Dov and Cohen \cite{Ben-DavCohen07}, respectively).
The upper bound, which is achieved in carefully controlled conditions, has been found by
Pfenninger \cite{Pfenninger61} as high as $\mathrm{Re}_{L} \approx 1 \times 10^5$.
Recently, Avila et al. \cite{Avila_Hof11} have also determined $\mathrm{Re}_{L}=2.04
\times 10^3$ as the critical point for the onset of sustained turbulence. In the present
study this point can be associated with the point where the low and upper boundaries of
the transition range become close to each other; according to Fig. \ref{fig4}, it
corresponds to $\mathrm{Re}_{L} \approx 1.4 \times 10^3$ ($\alpha \approx
\beta_{\mathrm{max}} \approx 1.25$).

\subsection{Other Characteristic Properties of Turbulent Flows}
\label{sec:2.2}%
The model is also consistent with some other characteristic properties of turbulent
flows:

$\bullet$ The dissipation scale $\eta$ obeys the known equation in the Kolmogorov theory
of turbulence \cite{Kolmogorov41a}
\begin{equation}\label{eq8}
\eta/L \sim {\mathrm {Re}}_{L}^{-3/4}
\end{equation}
Indeed, similar to $(N/n)^{1/3}$ in Eq. (\ref{eq7a}), the quantity
$q_{i}^{1/3}=(N_{i}/n)^{1/3}$ can be considered to be the Reynolds number of $i$
structure, i.e. $\mathrm{Re}_{i}=q_{i}^{1/3}$. According to Kolmogorov
\cite{Kolmogorov41a}, $\mathrm{Re}_{i} \sim 1$ characterizes the dissipation scale
$\eta$. Let us rewrite $q_{i}=N_{i}/n$ as $(N_{i}/N)\times (N/n)$ and take into account
that $v_{i}/V \sim (l_{i}/L)^{1/3}$ [Eq. (\ref{eq3})], so that $N_{i}/N \sim
[v_{i}l_{i}/(VL)]^{3} \sim (l_{i}/L)^4$. Then,
\begin{equation}\label{eq9}
q_i \sim (N/n)(l_i/L)^4=\mathrm{Re}_{L}^3(l_i/L)^4
\end{equation}
and the equality $q_{i} \sim 1$ (with $l_i \equiv \eta$) leads to Eq. (\ref{eq8}). This,
in particular, suggests that the linear size of the elementary cell in physical space
could be determined by the dissipation scale $\eta$, and the characteristic size of the
corresponding elementary volume of phase space (a cube root of its volume) could be
calculated as $\eta v_{\eta}$, where $v_{\eta}$ is the velocity increment at the
dissipation scale $v_{\eta} \sim V{\mathrm {Re}}_{L}^{-1/4}$. In other words, the size of
the volume is the the product of the Kolmogorov length $\eta =(\nu^{3}/\epsilon)^{1/4}$
and velocity $v_{\eta}=(\nu \epsilon)^{1/4}$ microscales, where $\epsilon$ is the rate of
dissipation per unit mass \cite{LandauLifshitz_fluid_mech,Kolmogorov41a}.

\begin{figure}[h]
\centering
\resizebox{0.8\columnwidth}{!}{ \includegraphics* {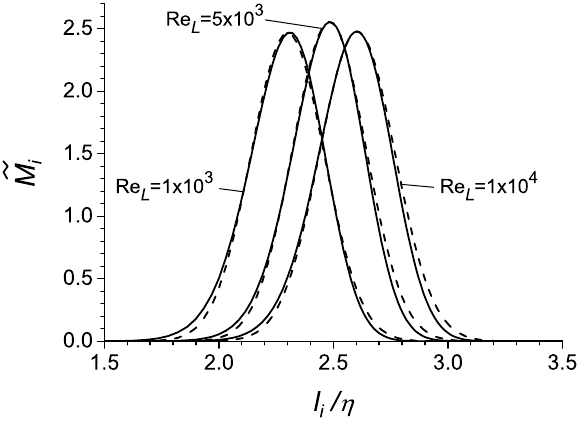}}%
\caption{Characteristic linear size distributions of the localized structures, $1 \leq
q_{i} \leq 200$. Solid lines show the $\tilde{M}_{i}(l_{i}/\eta)$ distributions averaged
over all combinations of $\alpha$ and $\beta$ for which $\mathrm {Re}_{L}$ are close to
those indicated at the curves within 1 percent (typically one of such distributions
dominated). The dashed curves are the Gaussian fits to the distributions.}
\label{fig5}       
\end{figure}

$\bullet$ The probability density distributions of the linear sizes of the structures
$\tilde{M}_{i}(l_{i}/\eta)$, Fig.\ref{fig5}, are in general agreement with the results of
the direct numerical simulation of isotropic homogeneous turbulence by Jim\'{e}nez {\it
{et al.}} \cite{JimenezWraySaffmanRogallo93}, who found that the vortex ("worm") radius
distributions are bell-shaped, practically do not shift with $\mathrm {Re}_{L}$ (with the
radius $r$ measured in $\eta$ units), and have maxima at $r \approx 3\eta$. According to
Eqs. (\ref{eq8}) and (\ref{eq9}),  $l_{i}/\eta \approx (q_i)^{1/4}$, so that the maxima
of the distributions at $l_{i}/\eta \approx 2.5$ correspond to $q_i \approx 40$.

$\bullet$ On the scale of localized structures, the structure functions and energy
spectrum are similar to those in the Kolmogorov-Obukhov theory
\cite{Kolmogorov41a,Obukhov41b}. Since the dependence $v=v(r)$ is assumed to be the same
for all localized structures, the structure functions are independent of the size
distributions of localized structures. With Eq. (\ref{eq3}), the structure function of
$p$-th order is $S_{p}(l)=\langle\delta v_{l}^{p}\rangle \sim l^{p/3}$, where $l$ is the
space increment, and the energy spectrum is $E(k) \sim k^{-5/3}$, where $k=2\pi/l$ is the
wave number.

\subsection{Another Type of Localized Structures}
\begin{figure}
\centering
\resizebox{0.8\columnwidth}{!}{ \includegraphics* {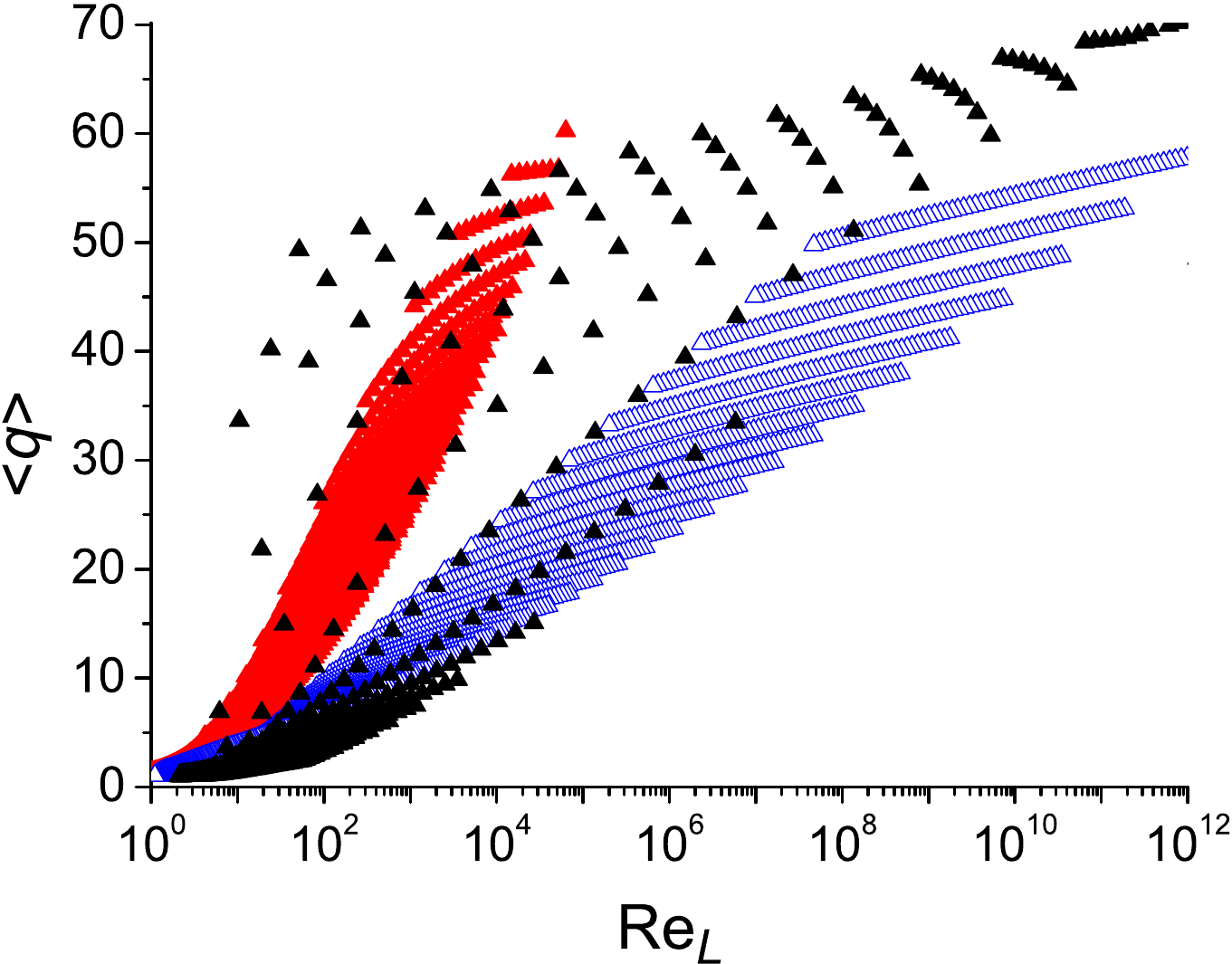}}%
\caption{Average number of the elementary cells in the localized structures $\langle q
\rangle$ as a function of $\mathrm{Re}_{L}$; $1 \leq q_{i} \leq 200$. The red triangles
are as those in Fig. \ref{eq3} ($-2.0 \leq \alpha \leq 2.0$ and $-2.0 \leq \beta \leq
2.0$), and the blue and black triangles correspond to the Burgers vortex with the core
radius $r_{\textrm c}=1$ and $r_{\textrm c}=7.5$, respectively (in both cases $-5.0 \leq
\alpha \leq 5.0$ and $-5.0 \leq \beta \leq 5.0$).}
\label{fig6}       
\end{figure}

The complete form of Eq. (\ref{eq3}), as it follows from the Kolmogorov dimensional
analysis \cite{Kolmogorov41a} (see also Landau and Lifshitz
\cite{LandauLifshitz_fluid_mech}), is $v \sim (\epsilon r)^{1/3}$. Therefore, it can not
be ruled out that the obtained behavior of $\langle q\rangle$ as a function of ${\mathrm
{Re}}_{L}$ (Figs. \ref{fig1} and \ref{fig3}) is a consequence of the fact that the
dynamical properties of the flow (through the dissipation rate $\epsilon$) are implicitly
taken into account. To see how the results change if the dependence of $v$ on $r$
considerably deviates from Eq. (\ref{eq3}), we performed calculations using the Burgers
vortex \cite{Burgers48} (see also Refs. \onlinecite{Saffman97,MarcuMeiburgNewton95}) to
mimic the localized structure. In this case the fluid motion varies from a solid-body
rotation within the vortex core ($v \sim r$) to the potential motion outside the core ($v
\sim 1/r$). The azimuthal velocity is written as $v \sim [1-\exp(-r/r_{\mathrm c})^2]/r$,
where $r_{\mathrm c}$ is the core radius. Then the dependence of $E_{i}$ on $N_{i}$ in
Eq. (\ref{eq2}) is determined by the equations $E_{i} \sim  l_z\int_{0}^{r_i} v^{2} r dr$
and $N_{i} \sim l_z r_{i}^{2}$, where $l_z$ is the vortex length; to make the vortices
somewhat similar to the "worms" of Ref. \onlinecite{JimenezWraySaffmanRogallo93}, we put
$l_z=5r_{\mathrm c}$. Figure \ref{fig6} compares the results for $r_{\mathrm c}=1$ and
$r_{\mathrm c}=7.5$ with what was obtained in the case of the Kolmogorov relation (Fig.
\ref{fig3}). The distributions at the above mentioned two values of $r_{\mathrm c}$ are
characteristically different is that in the former case the fluid motion is very close to
a potential motion for all $r$ ($v \sim 1/r$), while in the latter a pronounced
solid-body rotation exists at $r \lesssim r_{\mathrm c}$, where $v \sim r$. It is seen
that although the localized structures of the present type lead to less reasonable
results than those for the Kolmogorov dependence [Eq. (\ref{eq3})], at least for pipe
flows, the general tendency of increasing $\langle q\rangle$ with ${\mathrm {Re}}_{L}$ is
preserved.

\subsection{Connection with the Hydrodynamic Equations}
Summarizing the above given consideration (Sects. IIIA and IIIB), it can be said that the
proposed model is quite successful in description of characteristic properties of
turbulent flows, including the change of the flow regime with the Reynolds number. Then,
the natural question to ask is: Why do two apparently different approaches, the
hydrodynamic equations and the present model, lead to similar results? Or, what is common
to these approaches to lead to similar results? The present model is based on three
assumptions: i) the particles are identical, ii) there exists an elementary volume of
phase space in which the state of the particles is uncertain, and iii) the kinetic energy
of the fluctuations is as in the Kolmogorov theory for the inertial interval of scales.
According to the previous section, the latter assumption seems to be not essential for
our purpose - even if a considerable deviation from the Kolmogorov law is allowed, the
statistical preference of the structured flow is preserved (Fig. \ref{fig6}). The
attention should thus be focused on the former two assumptions. As well known, the
Navier-Stokes equation for a gas fluid can be derived from the Boltzmann equation (the
Chapman-Enskog expansion \cite{ChapmanCowling}), which, in turn, can be obtained from the
Liouville equation for Hamiltonian dynamics (see, e.g., Ferziger and Kaper
\cite{FerzigerKaper}). The transition from the Liouville equation to the Boltzmann
equation involves two essential assumptions. First, to reduce the many-particle
distribution function to the single-particle one  (the
Bogoliubov-Born-Green-Kirkwood-Yvon hierarchy; see, e.g., Kreuzer \cite{Kreuzer}), the
particles are taken to be identical. Secondly, because the gas density is considered to
be sufficiently low, the interaction between the particles is assumed to be a two-body
collision in which the states of the particles are uncorrelated prior to the collision
(the molecular chaos assumption \cite{FerzigerKaper}), so that the positions and
velocities of the particles are uncertain within the mean free path $\lambda$ and the
thermal velocity $c$, respectively. It thus follows that both the assumptions underlying
the present model are implicitly present in the Navier-Stokes equation \cite{liquid};
they can be said to be a hidden statistical content of the equation. Therefore, although
the hydrodynamic (Navier-Stokes) equations are incomparably rich in the description of
the turbulent phenomena, giving a dynamic picture of the transition and taking into
account the specific conditions under which the transition takes place (the inlet
conditions, flow environment, etc.), the driving force for the transition described with
the hydrodynamic equations can be of the same statistical origin as in the present model.

\section{Conclusions}
A model for turbulent fluid flow has been proposed that considers the flow as a
collection of localized spatial structures and estimates the statistical weights of these
collections for different Reynolds numbers. Each structure consists of elementary cells
in which the behavior of the particles is uncorrelated. To each such cell there
corresponds an elementary volume of phase space, in which the state of the particles is
uncertain. The Reynolds number is then associated with the ratio between the total phase
volume for the system and that for the elementary cell. It has been found that i) the
model successfully explains both the transition to turbulence at large Reynolds numbers
and characteristic properties of turbulent flows, and ii) the basic assumptions
underlying the model, i.e. that the particles are identical and an elementary volume of
the phase space exists where the state of the particles is uncertain, are implicitly
present in the Navier-Stokes equation. Considered together, these findings suggest that
with all the variety of dynamic phenomena that the hydrodynamic (Navier-Stokes) equations
describe, the driving force for turbulent transition is basically the same as in the
present model, i.e. the system tends to occupy a spatial state that is statistically
dominant at a given Reynolds number. The instability of the flow can then be a mechanism
to initiate the structural rearrangement of the flow to find this state.

The present statistical aspect of the turbulent transition can be important
for a more comprehensive understanding of the nature of this challenging phenomenon as well
as of the other related problems in the field of pattern formation \cite{CrossHohenberg93}.

\begin{acknowledgments}
I thank B. Ilyushin, Yu. Kachanov and D. Sikovsky for useful discussions and valuable
comments.
\end{acknowledgments}


\begin{references}
\bibitem{small_scales} At $r \ll \eta$, $v(r) \sim r$, as the first term of the expansion
of right-hand side of Eq. (3) in a series of powers of $r$; the constant at $r$ in this
equation is obtained by equating the right-hand sides of the given equation and Eq.
(\ref{eq3}) at $r=\eta$ (Landau and Lifshitz \cite{LandauLifshitz_fluid_mech}). In
principle, this dependence could be taking into account in Eq. (\ref{eq2}), extending the
lower bound of the range of variation of $N_i/n$ to $N_i/n \ll 1$. However, the effect of
the structures of such size is insignificant.
\bibitem{alpha_beta} The maximum and minimum values of $\alpha$ and $\beta$ were chosen as
follows. $\alpha_{\mathrm{max}}$ determined the upper boundary of the transition range in
terms of $\mathrm{Re}_{L}$ (Fig. \ref{fig3}). $\beta_{\mathrm{max}}$ was chosen such that
the lower boundary of transition was safely reached (Fig. \ref{fig3}).
$\alpha_{\mathrm{min}}$ and $\beta_{\mathrm{min}}$, which are negative, were chosen to be
sufficiently large in the absolute value not to affect the presented distributions.
\bibitem{liquid} A similar rigorous derivation of the Navier-Stokes equation for a liquid
fluid is problematic because of the many-body mechanism of interaction between the
particles \cite{Kreuzer}. However, the necessity of the assumptions that the particle are
identical and the system behaves stochastically on the microscale level seems to be
unavoidable.
\end{references}

\end{document}